\documentclass[sigconf,balance]{acmart}

\usepackage{booktabs} 
\usepackage{float}
\usepackage[utf8]{inputenc}
\usepackage{graphicx}
\usepackage{acronym}
\usepackage{longtable}
\usepackage{amsmath}
\usepackage{amssymb}
\usepackage{mathtools}
\usepackage[normalem]{ulem}
\usepackage{url}

\DeclareUrlCommand\ULurl{%
	\renewcommand\UrlLeft{\bgroup}%
	\renewcommand\UrlRight{\egroup}}

\setcopyright{rightsretained}

\acmYear{2019}
\copyrightyear{2019}

\begin{document}
\title{A Progressive Visual Analytics Tool for \\ Incremental Experimental Evaluation}

\author{\Large Fabio Giachelle \quad Gianmaria Silvello}
\author{\normalsize fabio.giachelle@unipd.it \quad gianmaria.silvello@unipd.it}

\affiliation{%
  \institution{Department of Information Engineering \\ University of Padua, Italy}
  \streetaddress{Via Giovanni Gradenigo, 6/b}
  \postcode{35131}
}

\renewcommand{\shortauthors}{F. Giachelle, G. Silvello}

\begin{abstract}
This paper presents a visual tool -- AVIATOR -- that integrates the \emph{progressive visual analytics} paradigm in the IR evaluation process. This tool serves to speed-up and facilitate the performance assessment of retrieval models enabling a result analysis through visual facilities. AVIATOR goes one  step beyond the common ``compute--wait--visualize'' analytics paradigm, introducing a continuous evaluation mechanism that minimizes human and computational resource consumption.
\end{abstract}


\maketitle

\section{Motivations}
The development of a new retrieval model is a demanding activity that goes beyond the definition and the implementation of the model itself. A retrieval model can be conceived as  part of an ecosystem where each component interacts with the others to produce the final document ranking for the user. As shown in~\cite{FerroSilvello2018}, the effectiveness of a model highly depends on the  pipeline components it interacts with (e.g., stoplist and stemmer). To determine which configuration is best in order to get the most out of a model is a demanding activity. In fact, it requires the inspection of several component pipelines and a comparison to baselines through multiple test collections and evaluation measures.  

The typical evaluation process comprises the following phases: corpus preprocessing (e.g., tokenization, stopword removal, stemming) and indexing phase, the retrieval phase and the evaluation phase itself. If something is modified in the preprocessing phase, the whole collection has to be re-indexed before testing the retrieval model and conducting the evaluation again. Unfortunately, indexing a collection may require hours, if not days, depending on the hardware and on the collection size. To assess the best configuration of components over multiple collections on the basis of a grid search requires great human effort and computational resources. 

We propose an ``all-in one visual analytics tool for the evaluation of IR systems'' (AVIATOR) to speed up this evaluation process. The idea behind the tool is to improve test retrieval models, calculate approximate measures, explore the results and make baseline comparisons during the indexing phase. AVIATOR allows the user to issue queries to a system while the indexing phase is still running and to explore partial evaluation results in an intuitive way thanks to visual analytics advances. In particular, leveraging on the \emph{progressive visual analytics paradigm} ``enable(s) an analyst to inspect partial results of an algorithm as they become available and interact with the algorithm to prioritize subspaces of interest''~\cite{StolperPG14}. 

Visual analytics and IR experimental evaluation have interacted before  producing visual tools to design  and ease failure analysis~\cite{AngeliniFSS14}, what-if analysis~\cite{AngeliniEtAl2016}, to explore pooling strategies~\cite{LipaniEtAl2017} and to enable interactive grid exploration over a large combinatorial space of systems~\cite{AngeliniFFSS18}. Nevertheless, they followed the ``compute-wait-visualize'' paradigm of visual analytics. AVIATOR moves a step beyond (partially) removing the ``wait'' phase. To the best of our knowledge, our paper is the first to focus on progressive visual analytics employed in IR to enable the dynamic and incremental evaluation of IR systems. 

A video showing the main functionalities of the system is available at the URL: \ULurl{https://www.gigasolution.it/v/Aviator.mp4}.

\textbf{Outline.} A general overview of AVIATOR is presented in Section \ref{sec:framework}. AVIATOR comprises of a back-end component that deals with the incremental indexing and retrieval (Section \ref{sec:backend}) and of a front-end component that enables the interactive exploration of the partial experimental results (Section \ref{sec:frontend}). 

\section{System Overview}\label{sec:framework}
AVIATOR embodies five phases: preprocessing, incremental indexing, retrieval, evaluation and visual analysis.  

In the preprocessing phase the document corpus $D$ is partitioned into $n$ bundles $\mathcal{B} = [B_1, B_2, \ldots, B_n] $, where $B_i, \text{ with } i \in [1,n-1]$, has size $k = \big\lfloor \frac{|D|}{n} \big\rfloor$ and $B_n$ has size $|D| - k(n-1)$. The bundles are populated by  uniformly sampling $D$ such that $B_i \cap B_j = \emptyset, \forall i,j \in [1,n]$. This sampling strategy is described in~\cite{HawkingR03}, where it is also shown that biased sub-collections exhibit similar behavior with uniform samples in terms of precision.

\begin{figure}[tbh]
\centering
\includegraphics[width=.9\columnwidth ]{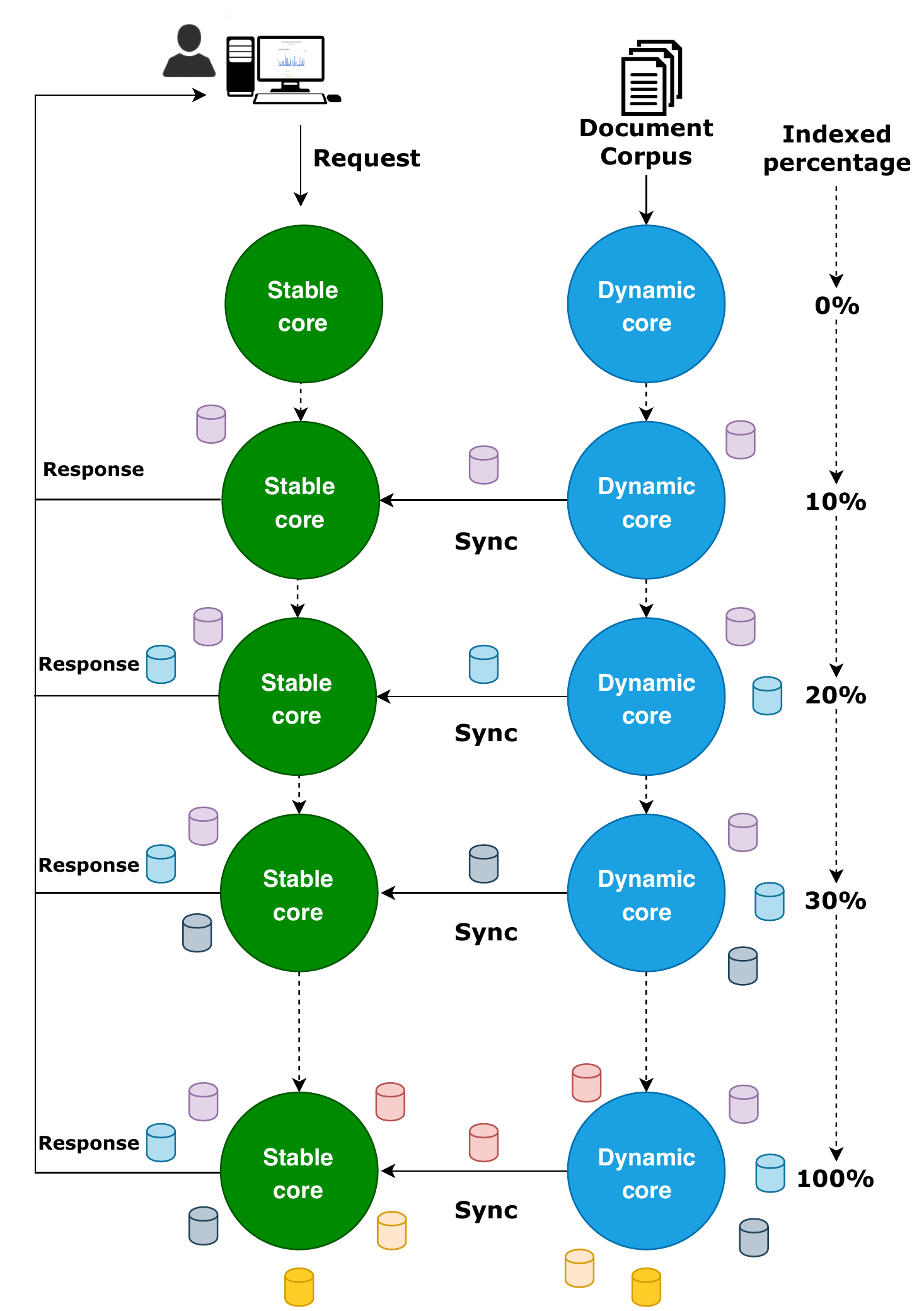}
\caption{Incremental indexing: the interaction between the stable and dynamic cores.}
    \label{figure:incremental_indexing}
\end{figure}

As shown in Figure \ref{figure:incremental_indexing}, in the incremental indexing phase, we adopt two parallel system threads each one implementing an independent instance of the same IRS. These threads are referred to as dynamic and stable core, respectively. The dynamic core indexes of the first corpus bundle and then releases the partial index to the stable core. The stable core  enables the user to run the retrieval phase on the partial index, while the dynamic core proceeds to index the second bundle. When the second bundle has been indexed, an interrupt is issued to the stable core and the user decides if s/he wants to update the index and run a new retrieval phase or to continue with the index already at hand.  

In the retrieval phase the partial index is queried by the user. Currently, AVIATOR is based on batch retrieval  on shared test collections. Hence, in each retrieval phase at least 50 queries are issued and a TREC-like run is returned for evaluation. The user can select several standard retrieval models or can use a custom one  loaded into the system. This phase can be considered as dynamic since the user can keep querying the partial index by changing the retrieval model or its parameters.

The runs produced in the retrieval phase undergo continuous evaluation as they are being produced. Once the evaluation phase is performed, the results are visualized by the visual analytics component that enables the user to conduct an in-depth and intuitive analysis. 

\section{Back-end component}\label{sec:backend}
The back-end component implements the first four phases described above. AVIATOR is a client-server application 
built on top of an IR system of choice. In the current implementation, AVIATOR is based on  Apache Solr~\footnote{\url{http://lucene.apache.org/solr/}} which in turn exploits the widely-used Apache Lucene search engine. In the back-end, AVIATOR acts as a wrapper on the IR system, controlling every stage of the IR process (indexing, retrieval and evaluation) via HTTP through a REpresentational State Transfer (REST)ful Web service. 

AVIATOR's demo version is based on the Disk 4\&5 of the TREC TIPSTER collection~\footnote{\url{https://trec.nist.gov/data/qa/T8_QAdata/disks4_5.html}} and on the $50$ topics (no. $351-400$) of the TREC7 ad-hoc track~\cite{VoorheesH98}. For testing purposes AVIATOR was designed to work with $64$ different IR system pipelines including four different stoplists (\texttt{indri, lucene, terrier, nostop}), four stemmers (\texttt{Hunspell, Krovetz, Porter, nostem}), and four IR models (\texttt{BM25, boolean\footnote{The \texttt{boolean} model, implemented in Apache Solr,  uses a simple matching coefficient to rank documents}, Dirichlet LM, TF-IDF}).

The incremental index is designed to  work on $10$ corpus bundles (10\%, 20\%, $\ldots$, 100\% of the corpus). This implies that, at the time of writing, the AVIATOR demo version works on 160 ($4 \times 4 \times 10$) different indexes that if statically stored in the memory would occupy up to $230$ GB. 

\begin{figure}[tbh]
\centering
\includegraphics[width=.9\columnwidth ]{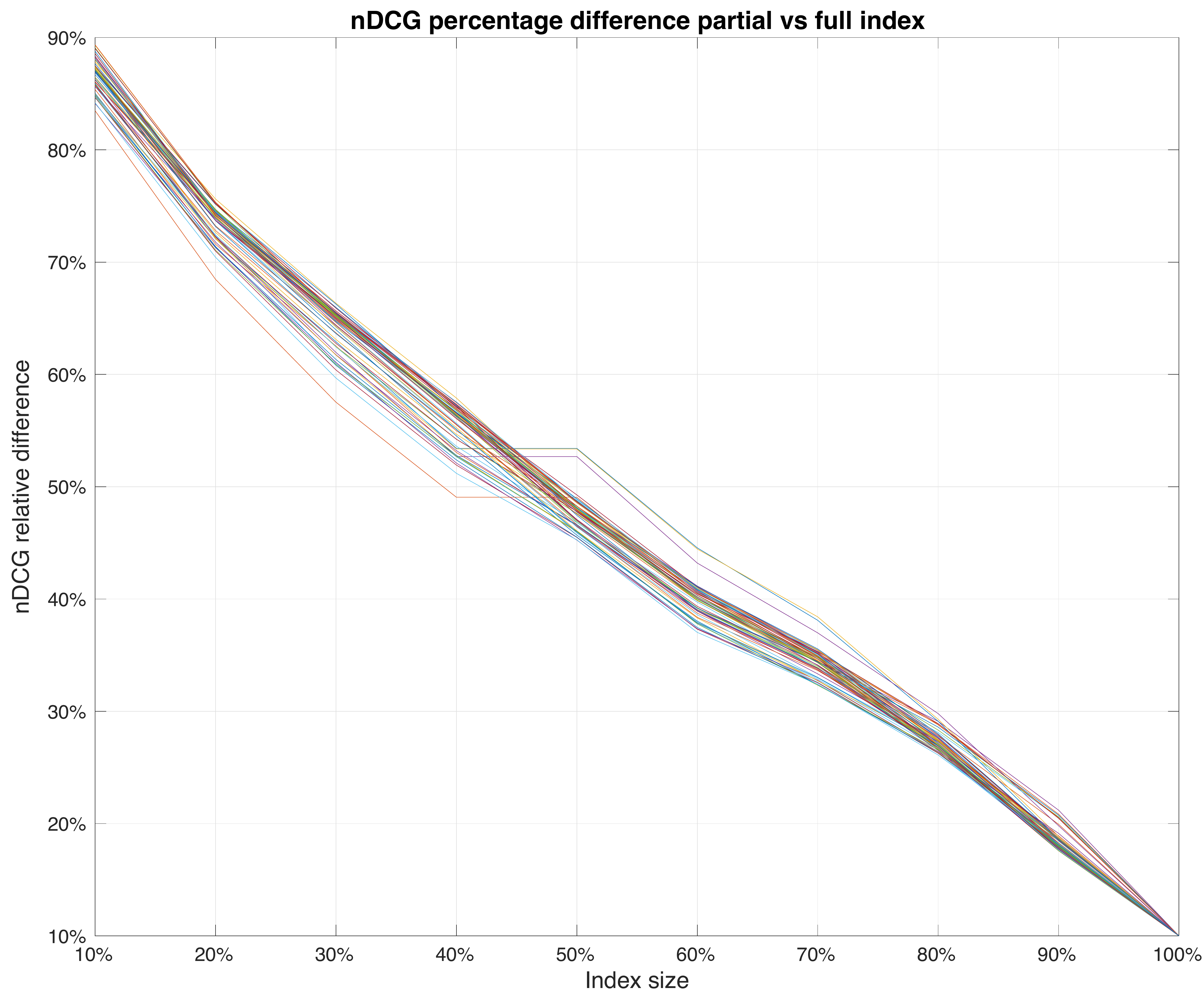}
\caption{Average nDCG relative difference between partial indexes at different levels of cut-off and the full index. Each line shows one of the 64 tested IR systems.}
    \label{figure:rel_diff}
\end{figure}

\begin{figure}[tbh]
\centering
\includegraphics[width=.9\columnwidth ]{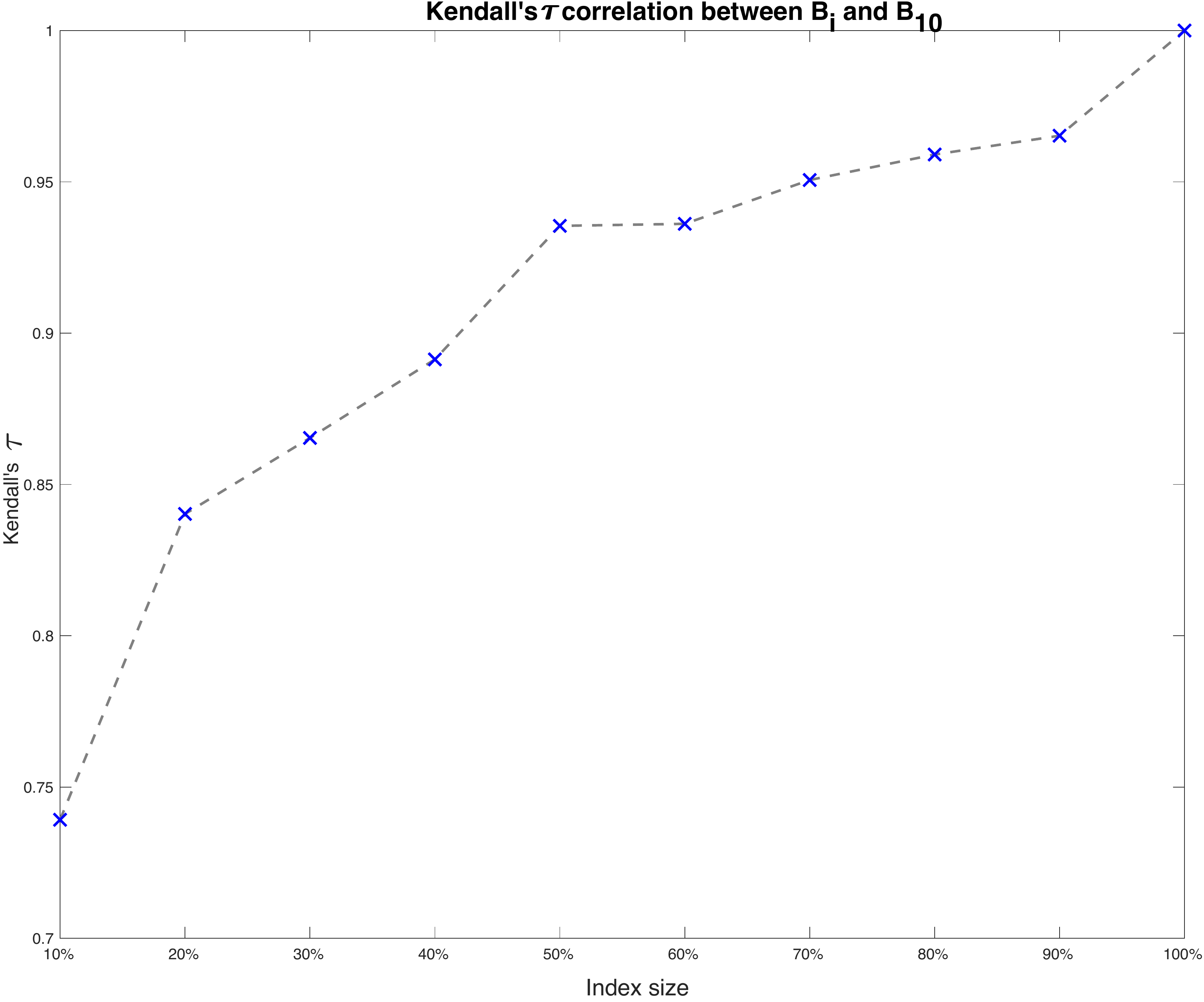}
\caption{Kendall's $\tau$ correlation between the system rankings (based on nDCG) obtained over increasingly more complete index bundles ($B_i, i \in [1,10]$) and the complete index bundle ($B_{10}$).}
    \label{figure:Kendall}
\end{figure}

The system run obtained over a partial index is an approximation of the ``true'' run obtained on the complete index. In Figure \ref{figure:rel_diff} we show the average nDCG was shown in relation to the system performance difference between  partial and full index. As expected, the precision of the measure grows with the index size and the approximate effectiveness is consistent across all the 64 tested systems. For instance, with index size $60\%$  for most of the systems the nDCG estimation is $40\%$ lower than the true value obtained with the full index. Figure \ref{figure:Kendall} shows that, on the TREC7, the system rankings obtained on partial indexes are highly correlated to the ranking obtained on the full one. The correlation is based on Kendall's $\tau$~\cite{Kendall48} and, following a common rule of thumb~\cite{Voorhees2001},  two rankings were found to be highly correlated, with $\tau>0.8$. Thus, in comparing all the 64 IR systems on the $20\%$ of the full index ($B_2$) our systems ranking is quite close to the one obtained with the full index. The correlation increases rather rapidly and when half of the collection is indexed, AVIATOR generates a reliable estimation of average system performances. 

\begin{figure}[tbh]
\centering
\includegraphics[width=.9\columnwidth ]{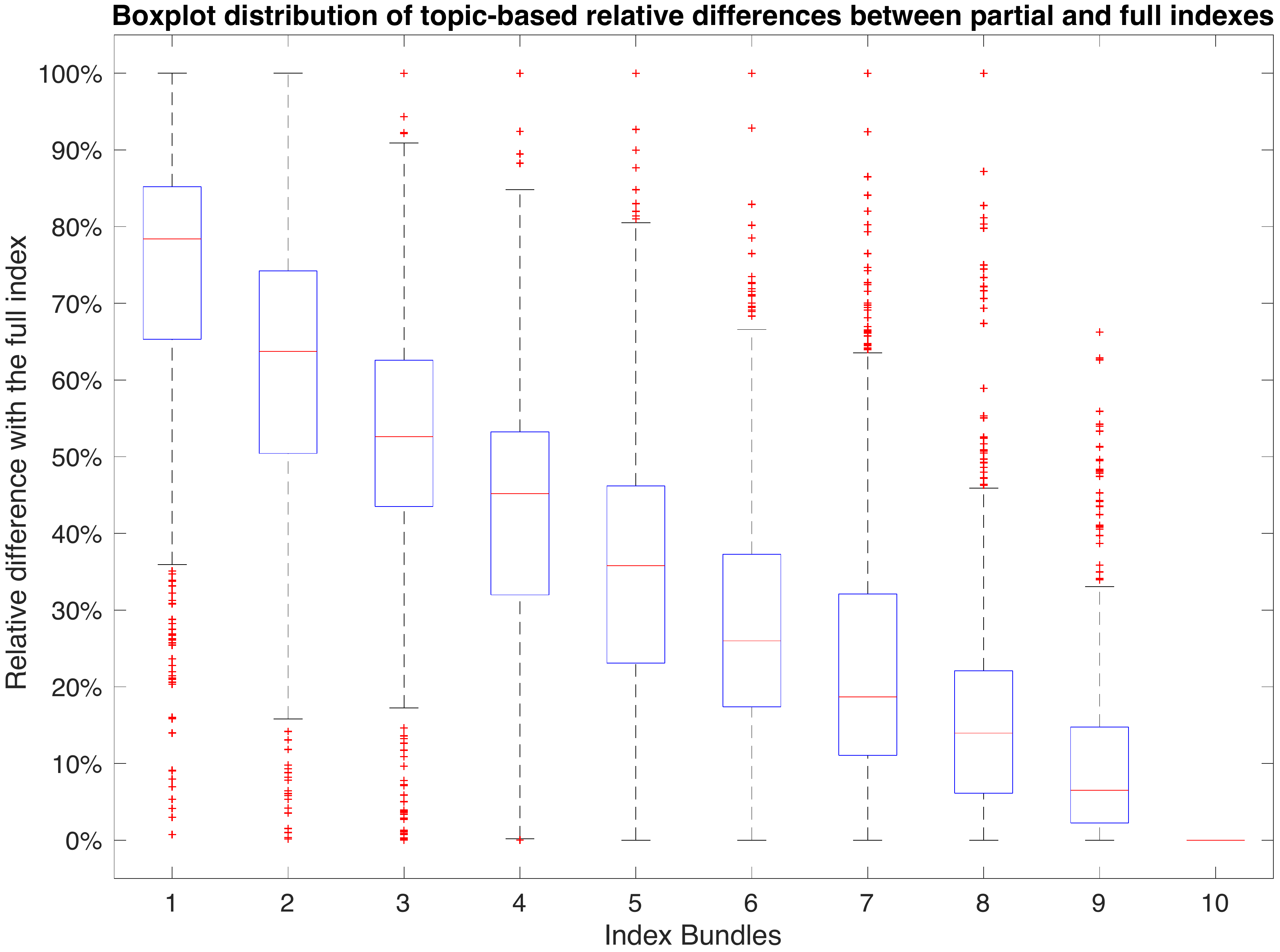}
\caption{Boxplot distribution of topic-based nDCG relative differences between partial and full indexes.}
    \label{figure:boxplot}
\end{figure}

Figure \ref{figure:boxplot} illustrates a topic based analysis of the nDCG where relative difference between the 64 runs are calculated on partial indexes and the final runs are calculated on the full index. We can see that with a $10\%$ index (bundle 1), half of the topics have an nDCG presenting a $80\%$ difference with the true nDCG value. Nevertheless, as can be seen, nDCG approximation improves steadily as the index grows. With half of the collection indexed (bundle 5) the nDCG approximation for half of the topics shows less than a $40\%$ difference from the final value.

\section{Front-end component}\label{sec:frontend}
The front-end component is a Web application designed on the basis of the Model-View-Controller design pattern. Its development  leverages on HTML5, D3~\footnote{\url{https://d3js.org}} and JQuery~\footnote{\url{https://jquery.com}} JavaScript libraries.  

\begin{figure}[tbh]
\centering
\includegraphics[width=1\columnwidth ]{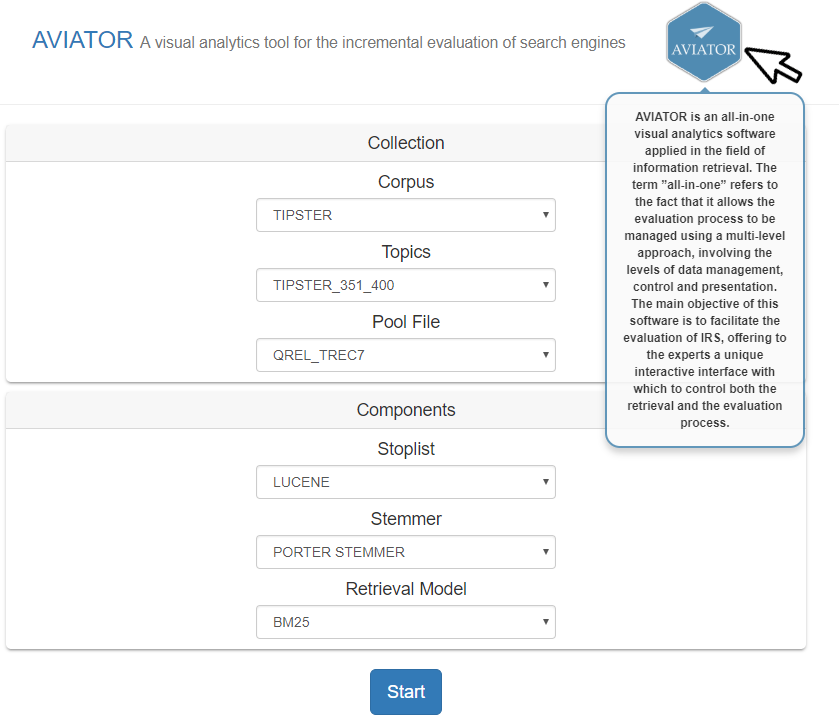}
\caption{The AVIATOR configuration interface.}
    \label{figure:configuration}
\end{figure}

Figure \ref{figure:configuration} shows the configuration page of AVIATOR. The user can select amongst different corpora, topic sets and pool files.  The current version of AVIATOR is based on the TIPSTER collection and TREC7 ad-hoc topics and pool file. For demo purposes the partial indexes have been precomputed. The  interaction with the system can therefore be artificially sped up to avoid the actual waiting time between one index version and the next. Moreover, the user can select the stoplist and the stemmer to be used for  building the index and a retrieval model. In turn, the retrieval model can be changed afterwards and other models can be added to the evaluation and analytics phase.

\begin{figure}[tbh]
\centering
\includegraphics[width=1\columnwidth ]{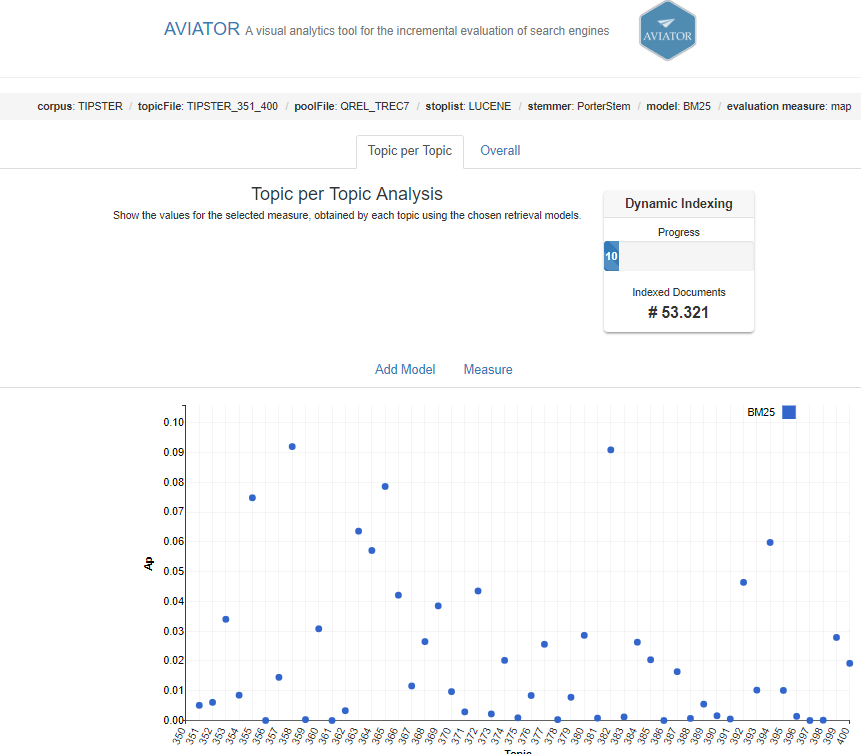}
\caption{The AVIATOR inspection interface: topic per topic visualization with a single model.}
    \label{figure:single_model_topic_per_topic}
\end{figure}

\begin{figure}[tbh]
\centering
\includegraphics[width=.8\columnwidth ]{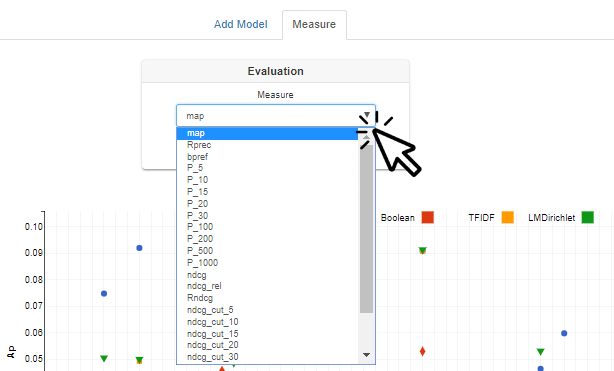}
\caption{The AVIATOR inspection interface: evaluation measure selection.}
    \label{figure:evaluation}
\end{figure}

Figure \ref{figure:single_model_topic_per_topic}  shows the main analytics interface for the topic based analysis. In the top of the screen, the main settings related to the collection and  index are reported, as a  reference for the user. Below, the two tabs can be used to conduct a topic based  or an overall analysis. In the top-right corner, the user can see the percentage of the corpus and the number of documents currently indexed. The main interaction interface shows a scatter plot with the Average Precision values of the retrieval model selected in the configuration phase. Just above the scatter plot, the two tabs can be used to add new retrieval models and to change the evaluation measure, as shown in Figure \ref{figure:evaluation} (all measures returned by \texttt{trec\_eval} are available).

\begin{figure}[tbh]
\centering
\includegraphics[width=1\columnwidth ]{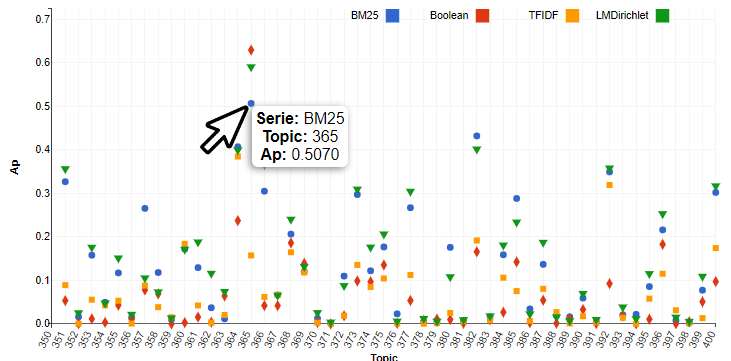}
\caption{The AVIATOR inspection interface: in-depth result analysis.}
    \label{figure:popup}
\end{figure}

Figure \ref{figure:popup} illustrates the scatter plot. Four different retrieval models can be compared through a pop-up window that can be triggered with a mouse over the points of the plot. The pop-up reports the retrieval model, the measure value and the topic being inspected. The user can zoom over a specific part of the scatter plot to better inspect the results. 

\begin{figure}[tbh]
\centering
\includegraphics[width=1\columnwidth ]{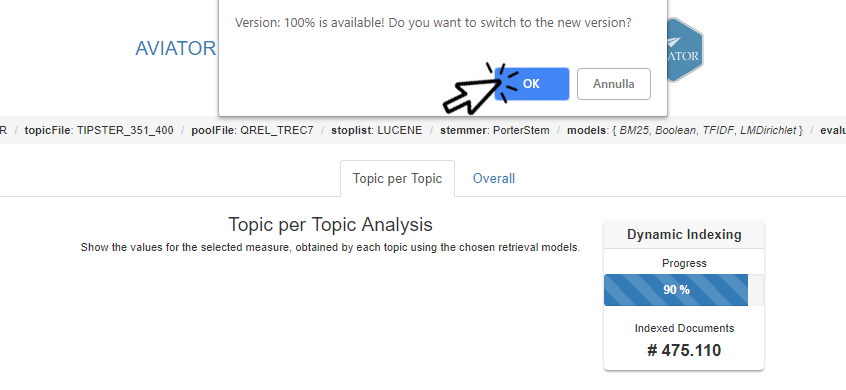}
\caption{The AVIATOR inspection interface: index update.}
    \label{figure:update}
\end{figure}

Figure \ref{figure:update} illustrates how the user is notified when a new version of the index is ready. The user can decide whether or not to work on a new version of the index. When a new version of the index is loaded all the visualizations are updated accordingly and the user settings are maintained from one version to the next.

\begin{figure}[tbh]
\centering
\includegraphics[width=1\columnwidth ]{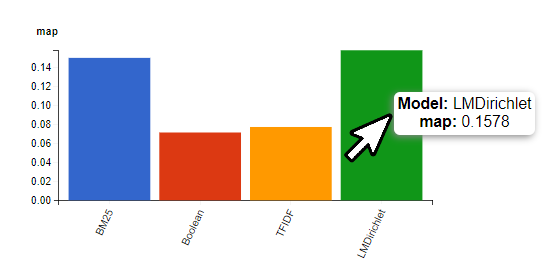}
\caption{The AVIATOR inspection interface: overall analysis.}
    \label{figure:overall}
\end{figure}

Figure \ref{figure:overall} shows the interface enabling the inspection of overall results (averaged over all topics) of the tested retrieval models. In this case too, a mouse over the plot bars triggers a pop-up window, providing detailed information on the inspected system.

\newpage

\section*{Diagrams, images, screen shots}

\newpage

\acrodef{3G}[3G]{Third Generation Mobile System}
\acrodef{5S}[5S]{Streams, Structures, Spaces, Scenarios, Societies}
\acrodef{AAAI}[AAAI]{Association for the Advancement of Artificial Intelligence}
\acrodef{AAL}[AAL]{Annotation Abstraction Layer}
\acrodef{AAM}[AAM]{Automatic Annotation Manager}
\acrodef{ACLIA}[ACLIA]{Advanced Cross-Lingual Information Access}
\acrodef{ACM}[ACM]{Association for Computing Machinery}
\acrodef{ADSL}[ADSL]{Asymmetric Digital Subscriber Line}
\acrodef{ADUI}[ADUI]{ADministrator User Interface}
\acrodef{AIP}[AIP]{Archival Information Package}
\acrodef{AJAX}[AJAX]{Asynchronous JavaScript Technology and \acs{XML}}
\acrodef{ALU}[ALU]{Aritmetic-Logic Unit}
\acrodef{AMUSID}[AMUSID]{Adaptive MUSeological IDentity-service}
\acrodef{ANOVA}[ANOVA]{ANalysis Of VAriance}
\acrodef{ANSI}[ANSI]{American National Standards Institute}
\acrodef{AP}[AP]{Average Precision}
\acrodef{APC}[APC]{AP Correlation}
\acrodef{API}[API]{Application Program Interface}
\acrodef{AR}[AR]{Address Register}
\acrodef{AS}[AS]{Annotation Service}
\acrodef{ASAP}[ASAP]{Adaptable Software Architecture Performance}
\acrodef{ASI}[ASI]{Annotation Service Integrator}
\acrodef{ASM}[ASM]{Annotation Storing Manager}
\acrodef{ASR}[ASR]{Automatic Speech Recognition}
\acrodef{ASUI}[ASUI]{ASsessor User Interface}
\acrodef{ATIM}[ATIM]{Annotation Textual Indexing Manager}
\acrodef{AUC}[AUC]{Area Under the ROC Curve}
\acrodef{AUI}[AUI]{Administrative User Interface}
\acrodef{AWARE}[AWARE]{Assessor-driven Weighted Averages for Retrieval Evaluation}
\acrodef{BANKS-I}[BANKS-I]{Browsing ANd Keyword Searching I}
\acrodef{BANKS-II}[BANKS-II]{Browsing ANd Keyword Searching II}
\acrodef{bpref}[bpref]{Binary Preference}
\acrodef{BNF}[BNF]{Backus and Naur Form}
\acrodef{BRICKS}[BRICKS]{Building Resources for Integrated Cultural Knowledge Services}
\acrodef{CAN}[CAN]{Content Addressable Netword}
\acrodef{CAS}[CAS]{Content-And-Structure}
\acrodef{CBSD}[CBSD]{Component-Based Software Developlement}
\acrodef{CBSE}[CBSE]{Component-Based Software Engineering}
\acrodef{CB-SPE}[CB-SPE]{Component-Based \acs{SPE}}
\acrodef{CD}[CD]{Collaboration Diagram}
\acrodef{CD}[CD]{Compact Disk}
\acrodef{CENL}[CENL]{Conference of European National Librarians}
\acrodef{CIDOC CRM}[CIDOC CRM]{CIDOC Conceptual Reference Model}
\acrodef{CIR}[CIR]{Current Instruction Register}
\acrodef{CIRCO}[CIRCO]{Coordinated Information Retrieval Components Orchestration}
\acrodef{CG}[CG]{Cumulated Gain}
\acrodef{CLAIRE}[CLAIRE]{Combinatorial visuaL Analytics system for Information Retrieval Evaluation}
\acrodef{CLEF}[CLEF]{Conference and Labs of the Evaluation Forum}
\acrodef{CLIR}[CLIR]{Cross Language Information Retrieval}
\acrodef{CMS}[CMS]{Content Management System}
\acrodef{CMT}[CMT]{Campaign Management Tool}
\acrodef{CNR}[CNR]{Italian National Council of Research}
\acrodef{CO}[CO]{Content-Only}
\acrodef{COD}[COD]{Code On Demand}
\acrodef{CODATA}[CODATA]{Committee on Data for Science and Technology}
\acrodef{COLLATE}[COLLATE]{Collaboratory for Annotation Indexing and Retrieval of Digitized Historical Archive Material}
\acrodef{CP}[CP]{Characteristic Pattern}
\acrodef{CPE}[CPE]{Control Processor Element}
\acrodef{CPU}[CPU]{Central Processing Unit}
\acrodef{CQL}[CQL]{Contextual Query Language}
\acrodef{CRP}[CRP]{Cumulated Relative Position}
\acrodef{CRUD}[CRUD]{Create--Read--Update--Delete}
\acrodef{CS}[CS]{Characteristic Structure}
\acrodef{CSM}[CSM]{Campaign Storing Manager}
\acrodef{CSS}[CSS]{Cascading Style Sheets}
\acrodef{CU}[CU]{Control Unit}
\acrodef{CUI}[CUI]{Client User Interface}
\acrodef{CV}[CV]{Cross-Validation}
\acrodef{DAFFODIL}[DAFFODIL]{Distributed Agents for User-Friendly Access of Digital Libraries}
\acrodef{DAO}[DAO]{Data Access Object}
\acrodef{DARE}[DARE]{Drawing Adequate REpresentations}
\acrodef{DARPA}[DARPA]{Defense Advanced Research Projects Agency}
\acrodef{DAS}[DAS]{Distributed Annotation System}
\acrodef{DB}[DB]{DataBase}
\acrodef{DBMS}[DBMS]{DataBase Management System}
\acrodef{DC}[DC]{Dublin Core}
\acrodef{DCG}[DCG]{Discounted Cumulated Gain}
\acrodef{DCMI}[DCMI]{Dublin Core Metadata Initiative}
\acrodef{DCV}[DCV]{Document Cut--off Value}
\acrodef{DD}[DD]{Deployment Diagram}
\acrodef{DDC}[DDC]{Dewey Decimal Classification}
\acrodef{DDS}[DDS]{Direct Data Structure}
\acrodef{DF}[DF]{Degrees of Freedom}
\acrodef{DFR}[DFR]{Divergence From Randomness}
\acrodef{DHT}[DHT]{Distributed Hash Table}
\acrodef{DI}[DI]{Digital Image}
\acrodef{DIKW}[DIKW]{Data, Information, Knowledge, Wisdom}
\acrodef{DIL}[DIL]{\acs{DIRECT} Integration Layer}
\acrodef{DiLAS}[DiLAS]{Digital Library Annotation Service}
\acrodef{DIRECT}[DIRECT]{Distributed Information Retrieval Evaluation Campaign Tool}
\acrodef{DKMS}[DKMS]{Data and Knowledge Management System}
\acrodef{DL}[DL]{Digital Library}
\acrodefplural{DL}[DL]{Digital Libraries}
\acrodef{DLMS}[DLMS]{Digital Library Management System}
\acrodef{DLOG}[DL]{Description Logics}
\acrodef{DLS}[DLS]{Digital Library System}
\acrodef{DLSS}[DLSS]{Digital Library Service System}
\acrodef{DO}[DO]{Digital Object}
\acrodef{DOI}[DOI]{Digital Object Identifier}
\acrodef{DOM}[DOM]{Document Object Model}
\acrodef{DoMDL}[DoMDL]{Document Model for Digital Libraries}
\acrodef{DPBF}[DPBF]{Dynamic Programming Best-First}
\acrodef{DR}[DR]{Data Register}
\acrodef{DRMM}[DRMM]{Deep Relevance Matching Model}
\acrodef{DRIVER}[DRIVER]{Digital Repository Infrastructure Vision for European Research}
\acrodef{DTD}[DTD]{Document Type Definition}
\acrodef{DVD}[DVD]{Digital Versatile Disk}
\acrodef{EAC-CPF}[EAC-CPF]{Encoded Archival Context for Corporate Bodies, Persons, and Families}
\acrodef{EAD}[EAD]{Encoded Archival Description}
\acrodef{EAN}[EAN]{International Article Number}
\acrodef{ECD}[ECD]{Enhanced Contenty Delivery}
\acrodef{ECDL}[ECDL]{European Conference on Research and Advanced Technology for Digital Libraries}
\acrodef{ECIR}[ECIR]{European Conference on Information Retrieval}
\acrodef{EDM}[EDM]{Europeana Data Model}
\acrodef{EG}[EG]{Execution Graph}
\acrodef{ELDA}[ELDA]{Evaluation and Language resources Distribution Agency}
\acrodef{ELRA}[ELRA]{European Language Resources Association}
\acrodef{EM}[EM]{Expectation Maximization}
\acrodef{EMMA}[EMMA]{Extensible MultiModal Annotation}
\acrodef{EPROM}[EPROM]{Erasable Programmable \acs{ROM}}
\acrodef{EQNM}[EQNM]{Extended Queueing Network Model}
\acrodef{ER}[ER]{Entity--Relationship}
\acrodef{ERR}[ERR]{Expected Reciprocal Rank}
\acrodef{ETL}[ETL]{Extract-Transform-Load}
\acrodef{FAST}[FAST]{Flexible Annotation Service Tool}
\acrodef{FIFO}[FIFO]{First-In / First-Out}
\acrodef{FIRE}[FIRE]{Forum for Information Retrieval Evaluation}
\acrodef{FN}[FN]{False Negative}
\acrodef{FNR}[FNR]{False Negative Rate}
\acrodef{FOAF}[FOAF]{Friend of a Friend}
\acrodef{FORESEE}[FORESEE]{FOod REcommentation sErvER}
\acrodef{FP}[FP]{False Positive}
\acrodef{FPR}[FPR]{False Positive Rate}
\acrodef{GIF}[GIF]{Graphics Interchange Format}
\acrodef{GIR}[GIR]{Geografic Information Retrieval}
\acrodef{GAP}[GAP]{Graded Average Precision}
\acrodef{GLM}[GLM]{General Linear Model}
\acrodef{GLMM}[GLMM]{General Linear Mixed Model}
\acrodef{GMAP}[GMAP]{Geometric Mean Average Precision}
\acrodef{GoP}[GoP]{Grid of Points}
\acrodef{GPRS}[GPRS]{General Packet Radio Service}
\acrodef{gRBP}[gRBP]{Graded Rank-Biased Precision}
\acrodef{GTIN}[GTIN]{Global Trade Item Number}
\acrodef{GUI}[GUI]{Graphical User Interface}
\acrodef{GW}[GW]{Gateway}
\acrodef{HCI}[HCI]{Human Computer Interaction}
\acrodef{HDS}[HDS]{Hybrid Data Structure}
\acrodef{HIR}[HIR]{Hypertext Information Retrieval}
\acrodef{HIT}[HIT]{Human Intelligent Task}
\acrodef{HITS}[HITS]{Hyperlink-Induced Topic Search}
\acrodef{HTML}[HTML]{HyperText Markup Language}
\acrodef{HTTP}[HTTP]{HyperText Transfer Protocol}
\acrodef{HSD}[HSD]{Honestly Significant Difference}
\acrodef{ICA}[ICA]{International Council on Archives}
\acrodef{ICSU}[ICSU]{International Council for Science}
\acrodef{IDF}[IDF]{Inverse Document Frequency}
\acrodef{IDS}[IDS]{Inverse Data Structure}
\acrodef{IEEE}[IEEE]{Institute of Electrical and Electronics Engineers}
\acrodef{IEI}[IEI]{Istituto della Enciclopedia Italiana fondata da Giovanni Treccani}
\acrodef{IETF}[IETF]{Internet Engineering Task Force}
\acrodef{IMS}[IMS]{Information Management System}
\acrodef{IMSPD}[IMS]{Information Management Systems Research Group}
\acrodef{indAP}[indAP]{Induced Average Precision}
\acrodef{infAP}[infAP]{Inferred Average Precision}
\acrodef{INEX}[INEX]{INitiative for the Evaluation of \acs{XML} Retrieval}
\acrodef{INS-M}[INS-M]{Inverse Set Data Model}
\acrodef{INTR}[INTR]{Interrupt Register}
\acrodef{IP}[IP]{Internet Protocol}
\acrodef{IPSA}[IPSA]{Imaginum Patavinae Scientiae Archivum}
\acrodef{IR}[IR]{Information Retrieval}
\acrodef{IRON}[IRON]{Information Retrieval ON}
\acrodef{IRON2}[IRON$^2$]{Information Retrieval On aNNotations}
\acrodef{IRON-SAT}[IRON-SAT]{\acs{IRON} - Statistical Analysis Tool}
\acrodef{IRS}[IRS]{Information Retrieval System}
\acrodef{ISAD(G)}[ISAD(G)]{International Standard for Archival Description (General)}
\acrodef{ISBN}[ISBN]{International Standard Book Number}
\acrodef{ISIS}[ISIS]{Interactive SImilarity Search}
\acrodef{ISJ}[ISJ]{Interactive Searching and Judging}
\acrodef{ISO}[ISO]{International Organization for Standardization}
\acrodef{ITU}[ITU]{International Telecommunication Union }
\acrodef{ITU-T}[ITU-T]{Telecommunication Standardization Sector of \acs{ITU}}
\acrodef{IV}[IV]{Information Visualization}
\acrodef{JAN}[JAN]{Japanese Article Number}
\acrodef{JDBC}[JDBC]{Java DataBase Connectivity}
\acrodef{JMB}[JMB]{Java--Matlab Bridge}
\acrodef{JPEG}[JPEG]{Joint Photographic Experts Group}
\acrodef{JSON}[JSON]{JavaScript Object Notation}
\acrodef{JSP}[JSP]{Java Server Pages}
\acrodef{JTE}[JTE]{Java-Treceval Engine}
\acrodef{KDE}[KDE]{Kernel Density Estimation}
\acrodef{KLD}[KLD]{Kullback-Leibler Divergence}
\acrodef{KLAPER}[KLAPER]{Kernel LAnguage for PErformance and Reliability analysis}
\acrodef{LAM}[LAM]{Libraries, Archives, and Museums}
\acrodef{LAM2}[LAM]{Logistic Average Misclassification}
\acrodef{LAN}[LAN]{Local Area Network}
\acrodef{LD}[LD]{Linked Data}
\acrodef{LEAF}[LEAF]{Linking and Exploring Authority Files}
\acrodef{LIDO}[LIDO]{Lightweight Information Describing Objects}
\acrodef{LIFO}[LIFO]{Last-In / First-Out}
\acrodef{LM}[LM]{Language Model}
\acrodef{LMT}[LMT]{Log Management Tool}
\acrodef{LOD}[LOD]{Linked Open Data}
\acrodef{LODE}[LODE]{Linking Open Descriptions of Events}
\acrodef{LpO}[LpO]{Leave-$p$-Out}
\acrodef{LRM}[LRM]{Local Relational Model}
\acrodef{LRU}[LRU]{Last Recently Used}
\acrodef{LS}[LS]{Lexical Signature}
\acrodef{LSM}[LSM]{Log Storing Manager}
\acrodef{LUG}[LUG]{Lexical Unit Generator}
\acrodef{MA}[MA]{Mobile Agent}
\acrodef{MA}[MA]{Moving Average}
\acrodef{MACS}[MACS]{Multilingual ACcess to Subjects}
\acrodef{MADCOW}[MADCOW]{Multimedia Annotation of Digital Content Over the Web}
\acrodef{MAD}[MAD]{Mean Assessed Documents}
\acrodef{MADP}[MADP]{Mean Assessed Documents Precision}
\acrodef{MADS}[MADS]{Metadata Authority Description Standard}
\acrodef{MAP}[MAP]{Mean Average Precision}
\acrodef{MARC}[MARC]{Machine Readable Cataloging}
\acrodef{MATTERS}[MATTERS]{MATlab Toolkit for Evaluation of information Retrieval Systems}
\acrodef{MDA}[MDA]{Model Driven Architecture}
\acrodef{MDD}[MDD]{Model-Driven Development}
\acrodef{METS}[METS]{Metadata Encoding and Transmission Standard}
\acrodef{MIDI}[MIDI]{Musical Instrument Digital Interface}
\acrodef{MIME}[MIME]{Multipurpose Internet Mail Extensions}
\acrodef{MLIA}[MLIA]{MultiLingual Information Access}
\acrodef{MM}[MM]{Machinery Model}
\acrodef{MMU}[MMU]{Memory Management Unit}
\acrodef{MODS}[MODS]{Metadata Object Description Schema}
\acrodef{MOF}[MOF]{Meta-Object Facility}
\acrodef{MP}[MP]{Markov Precision}
\acrodef{MPEG}[MPEG]{Motion Picture Experts Group}
\acrodef{MRD}[MRD]{Machine Readable Dictionary}
\acrodef{MRF}[MRF]{Markov Random Field}
\acrodef{MS}[MS]{Mean Squares}
\acrodef{MSAC}[MSAC]{Multilingual Subject Access to Catalogues}
\acrodef{MSE}[MSE]{Mean Square Error}
\acrodef{MT}[MT]{Machine Translation}
\acrodef{MV}[MV]{Majority Vote}
\acrodef{MVC}[MVC]{Model-View-Controller}
\acrodef{NACSIS}[NACSIS]{NAtional Center for Science Information Systems}
\acrodef{NAP}[NAP]{Network processors Applications Profile}
\acrodef{NCP}[NCP]{Normalized Cumulative Precision}
\acrodef{nCG}[nCG]{Normalized Cumulated Gain}
\acrodef{nCRP}[nCRP]{Normalized Cumulated Relative Position}
\acrodef{nDCG}[nDCG]{Normalized Discounted Cumulated Gain}
\acrodef{NESTOR}[NESTOR]{NEsted SeTs for Object hieRarchies}
\acrodef{NeuIR}[NeuIR]{Neural Information Retrieval}
\acrodef{NEXI}[NEXI]{Narrowed Extended XPath I}
\acrodef{NII}[NII]{National Institute of Informatics}
\acrodef{NISO}[NISO]{National Information Standards Organization}
\acrodef{NIST}[NIST]{National Institute of Standards and Technology}
\acrodef{NLP}[NLP]{Natural Language Processing}
\acrodef{NP}[NP]{Network Processor}
\acrodef{NR}[NR]{Normalized Recall}
\acrodef{NS-M}[NS-M]{Nested Set Model}
\acrodef{NTCIR}[NTCIR]{NII Testbeds and Community for Information access Research}
\acrodef{NVSM}[NVSM]{Neural Vector Space Model}
\acrodef{OAI}[OAI]{Open Archives Initiative}
\acrodef{OAI-ORE}[OAI-ORE]{Open Archives Initiative Object Reuse and Exchange}
\acrodef{OAI-PMH}[OAI-PMH]{Open Archives Initiative Protocol for Metadata Harvesting}
\acrodef{OAIS}[OAIS]{Open Archival Information System}
\acrodef{OC}[OC]{Operation Code}
\acrodef{OCLC}[OCLC]{Online Computer Library Center}
\acrodef{OMG}[OMG]{Object Management Group}
\acrodef{OLAP}[OLAP]{On-Line Analytical Processing}
\acrodef{OO}[OO]{Object Oriented}
\acrodef{OODB}[OODB]{Object-Oriented \acs{DB}}
\acrodef{OODBMS}[OODBMS]{Object-Oriented \acs{DBMS}}
\acrodef{OPAC}[OPAC]{Online Public Access Catalog}
\acrodef{OQL}[OQL]{Object Query Language}
\acrodef{ORP}[ORP]{Open Relevance Project}
\acrodef{OSIRIS}[OSIRIS]{Open Service Infrastructure for Reliable and Integrated process Support}
\acrodef{P2P}[P2P]{Peer-To-Peer}
\acrodef{PA}[PA]{Performance Analysis}
\acrodef{PAMT}[PAMT]{Pool-Assessment Management Tool}
\acrodef{PASM}[PASM]{Pool-Assessment Storing Manager}
\acrodef{PC}[PC]{Program Counter}
\acrodef{PCP}[PCP]{Pre-Commercial Procurement}
\acrodef{PCR}[PCR]{Peripherical Command Register}
\acrodef{PDA}[PDA]{Personal Digital Assistant}
\acrodef{PDF}[PDF]{Probability Density Function}
\acrodef{PDR}[PDR]{Peripherical Data Register}
\acrodef{POI}[POI]{\acs{PURL}-based Object Identifier}
\acrodef{PoS}[PoS]{Part of Speech}
\acrodef{PPE}[PPE]{Programmable Processing Engine}
\acrodef{PREFORMA}[PREFORMA]{PREservation FORMAts for culture information/e-archives}
\acrodef{PRIMAmob-UML}[PRIMAmob-UML]{mobile \acs{PRIMA-UML}}
\acrodef{PRIMA-UML}[PRIMA-UML]{PeRformance IncreMental vAlidation in \acs{UML}}
\acrodef{PROM}[PROM]{Programmable \acs{ROM}}
\acrodef{PROMISE}[PROMISE]{Participative Research labOratory  for Multimedia and Multilingual Information Systems Evaluation}
\acrodef{pSQL}[pSQL]{propagate \acs{SQL}}
\acrodef{PUI}[PUI]{Participant User Interface}
\acrodef{PURL}[PURL]{Persistent \acs{URL}}
\acrodef{QA}[QA]{Question Answering}
\acrodef{QoS-UML}[QoS-UML]{\acs{UML} Profile for QoS and Fault Tolerance}
\acrodef{RAM}[RAM]{Random Access Memory}
\acrodef{RAMM}[RAM]{Random Access Machine}
\acrodef{RBO}[RBO]{Rank-Biased Overlap}
\acrodef{RBP}[RBP]{Rank-Biased Precision}
\acrodef{RDBMS}[RDBMS]{Relational \acs{DBMS}}
\acrodef{RDF}[RDF]{Resource Description Framework}
\acrodef{REST}[REST]{REpresentational State Transfer}
\acrodef{REV}[REV]{Remote Evaluation}
\acrodef{RFC}[RFC]{Request for Comments}
\acrodef{RIA}[RIA]{Reliable Information Access}
\acrodef{RMSE}[RMSE]{Root Mean Square Error}
\acrodef{RMT}[RMT]{Run Management Tool}
\acrodef{ROM}[ROM]{Read Only Memory}
\acrodef{ROMIP}[ROMIP]{Russian Information Retrieval Evaluation Seminar}
\acrodef{RoMP}[RoMP]{Rankings of Measure Pairs}
\acrodef{RoS}[RoS]{Rankings of Systems}
\acrodef{RP}[RP]{Relative Position}
\acrodef{RR}[RR]{Reciprocal Rank}
\acrodef{RSM}[RSM]{Run Storing Manager}
\acrodef{RST}[RST]{Rhetorical Structure Theory}
\acrodef{RT-UML}[RT-UML]{\acs{UML} Profile for Schedulability, Performance and Time}
\acrodef{SA}[SA]{Software Architecture}
\acrodef{SAL}[SAL]{Storing Abstraction Layer}
\acrodef{SAMT}[SAMT]{Statistical Analysis Management Tool}
\acrodef{SAN}[SAN]{Sistema Archivistico Nazionale}
\acrodef{SASM}[SASM]{Statistical Analysis Storing Manager}
\acrodef{SD}[SD]{Sequence Diagram}
\acrodef{SE}[SE]{Search Engine}
\acrodef{SEBD}[SEBD]{Convegno Nazionale su Sistemi Evoluti per Basi di Dati}
\acrodef{SFT}[SFT]{Satisfaction--Frustration--Total}
\acrodef{SIGIR}[SIGIR]{ACM SIGIR Conference on Research \& Development in Information Retrieval}
\acrodef{SIL}[SIL]{Service Integration Layer}
\acrodef{SIP}[SIP]{Submission Information Package}
\acrodef{SKOS}[SKOS]{Simple Knowledge Organization System}
\acrodef{SM}[SM]{Software Model}
\acrodef{SMART}[SMART]{System for the Mechanical Analysis and Retrieval of Text}
\acrodef{SoA}[SoA]{Service-oriented Architectures}
\acrodef{SOA}[SOA]{Strength of Association}
\acrodef{SOAP}[SOAP]{Simple Object Access Protocol}
\acrodef{SOM}[SOM]{Self-Organizing Map}
\acrodef{SPE}[SPE]{Software Performance Engineering}
\acrodef{SPINA}[SPINA]{Superimposed Peer Infrastructure for iNformation Access}
\acrodef{SPLIT}[SPLIT]{Stemming Program for Language Independent Tasks}
\acrodef{SPOOL}[SPOOL]{Simultaneous Peripheral Operations On Line}
\acrodef{SQL}[SQL]{Structured Query Language}
\acrodef{SR}[SR]{Sliding Ratio}
\acrodef{SR}[SR]{Status Register}
\acrodef{SRU}[SRU]{Search/Retrieve via \acs{URL}}
\acrodef{SS}[SS]{Sum of Squares}
\acrodef{SSTF}[SSTF]{Shortest Seek Time First}
\acrodef{STAR}[STAR]{Steiner-Tree Approximation in Relationship graphs}
\acrodef{STON}[STON]{STemming ON}
\acrodef{TAC}[TAC]{Text Analysis Conference}
\acrodef{TBG}[TBG]{Time-Biased Gain}
\acrodef{TCP}[TCP]{Transmission Control Protocol}
\acrodef{TEL}[TEL]{The European Library}
\acrodef{TERRIER}[TERRIER]{TERabyte RetrIEveR}
\acrodef{TF}[TF]{Term Frequency}
\acrodef{TFR}[TFR]{True False Rate}
\acrodef{TN}[TN]{True Negative}
\acrodef{TO}[TO]{Transfer Object}
\acrodef{TP}[TP]{True Positve}
\acrodef{TPR}[TPR]{True Positive Rate}
\acrodef{TRAT}[TRAT]{Text Relevance Assessing Task}
\acrodef{TREC}[TREC]{Text REtrieval Conference}
\acrodef{TRECVID}[TRECVID]{TREC Video Retrieval Evaluation}
\acrodef{TTL}[TTL]{Time-To-Live}
\acrodef{UCD}[UCD]{Use Case Diagram}
\acrodef{UDC}[UDC]{Universal Decimal Classification}
\acrodef{uGAP}[uGAP]{User-oriented Graded Average Precision}
\acrodef{UI}[UI]{User Interface}
\acrodef{UML}[UML]{Unified Modeling Language}
\acrodef{UMT}[UMT]{User Management Tool}
\acrodef{UMTS}[UMTS]{Universal Mobile Telecommunication System}
\acrodef{UoM}[UoM]{Utility-oriented Measurement}
\acrodef{UPC}[UPC]{Universal Product Code}
\acrodef{URI}[URI]{Uniform Resource Identifier}
\acrodef{URL}[URL]{Uniform Resource Locator}
\acrodef{URN}[URN]{Uniform Resource Name}
\acrodef{USM}[USM]{User Storing Manager}
\acrodef{VA}[VA]{Visual Analytics}
\acrodef{VATE}[VATE$^2$]{Visual Analytics Tool for Experimental Evaluation}
\acrodef{VIRTUE}[VIRTUE]{Visual Information Retrieval Tool for Upfront Evaluation}
\acrodef{VD}[VD]{Virtual Document}
\acrodef{VIAF}[VIAF]{Virtual International Authority File}
\acrodef{VL}[VL]{Visual Language}
\acrodef{VoIP}[VoIP]{Voice over IP}
\acrodef{VS}[VS]{Visual Sentence}
\acrodef{W3C}[W3C]{World Wide Web Consortium}
\acrodef{WAN}[WAN]{Wide Area Network}
\acrodef{WHO}[WHO]{World Health Organization}
\acrodef{WLAN}[WLAN]{Wireless \acs{LAN}}
\acrodef{WP}[WP]{Work Package}
\acrodef{WS}[WS]{Web Services}
\acrodef{WSD}[WSD]{Word Sense Disambiguation}
\acrodef{WSDL}[WSDL]{Web Services Description Language}
\acrodef{WWW}[WWW]{World Wide Web}
\acrodef{XMI}[XMI]{\acs{XML} Metadata Interchange}
\acrodef{XML}[XML]{eXtensible Markup Language}
\acrodef{XPath}[XPath]{XML Path Language}
\acrodef{XSL}[XSL]{eXtensible Stylesheet Language}
\acrodef{XSL-FO}[XSL-FO]{\acs{XSL} Formatting Objects}
\acrodef{XSLT}[XSLT]{\acs{XSL} Transformations}
\acrodef{YAGO}[YAGO]{Yet Another Great Ontology}
\acrodef{YASS}[YASS]{Yet Another Suffix Stripper}

\bibliographystyle{ACM-Reference-Format}
\bibliography{bibliografia,zz-proceedings}

\end{document}